# An Interacting Binary System Powers Precessing Outflows of an Evolved Star

Henri M. J. Boffin[1*], Brent Miszalski[2,3], Thomas Rauch[4], David Jones[1], Romano L. M. Corradi[5,6], Ralf Napiwotzki[7], Avril C. Day-Jones[8], and Joachim Köppen[9]

[1]European Southern Observatory, Alonso de Cordova 3107, Vitacura, Casilla 19001, Santiago, Chile
[2]South African Astronomical Observatory, PO Box 9, Observatory 7935, South Africa
[3]Southern African Large Telescope Foundation, PO Box 9, Observatory 7935, South Africa
[4]Institute for Astronomy and Astrophysics, Kepler Center for Astro and Particle Physics, Eberhard Karls University, Sand 1, 72076 Tübingen, Germany
[5]Instituto de Astrofísica de Canarias, E-38200 La Laguna, Tenerife, Spain
[6]Departamento de Astrofísica, Universidad de La Laguna, E-38206 La Laguna, Tenerife, Spain
[7]University of Hertfordshire, Physics Astronomy and Mathematics, Hatfield AL10 9AB, UK
[8]Universidad de Chile, Santiago, Casilla 36-D, Chile
[9]Observatoire de Strasbourg, 11 rue de l'université, F-67000 Strasbourg, France

[*]To whom correspondence should be addressed. E-mail: hboffin@eso.org

**Stars are generally spherical, yet their gaseous envelopes often appear non-spherical when ejected near the end of their lives. This quirk is most notable during the planetary nebula phase when these envelopes become ionized. Interactions among stars in a binary system are suspected to cause the asymmetry. In particular, a precessing accretion disk around a companion is believed to launch point-symmetric jets, as seen in the prototype Fleming 1. Our discovery of a post common-envelope binary nucleus in Fleming 1 confirms that this scenario is highly favorable. Similar binary interactions are therefore likely to explain these kinds of outflows in a large variety of systems.**

Planetary nebulae (PNe) are thought to represent the transitory phase of the end of the lives of solar-like stars. The mass-loss mechanisms at play during the late stages of stellar evolution that produce the observed shapes of planetary nebulae have been a matter of debate in the last two decades (*1*). The leading paradigm to produce the most extreme nebular morphologies is evolution in an interacting binary system (*2-4*), in particular common-envelope (CE) evolution - the dramatic outcome of unstable mass transfer resulting in a binary system with a greatly reduced orbital period (P<~1 day for PNe). Despite recent detections of multiple post common-envelope binary central stars (*5-7*), there are as yet no clear-cut examples of binaries actively shaping their surrounding planetary nebulae. A handful of post-CE nebulae are known to be oriented in agreement with the orbital inclination of the binaries that ejected them (*8*) – as would be expected. However, we do not yet have any inkling how a particular binary configuration gives rise to a specific fundamental nebula shape. An alternative approach to tackle this difficult problem is to focus instead on nebular features that are impossible to explain without binaries. These are the low-ionisation microstructures that include filaments, knots and collimated fast outflows (hereafter jets). Theory prescribes that jets are launched and collimated by accretion, rotation and/or magnetic mechanisms (*9*), but in the case of PNe only the binary explanation is feasible (*10*). In fact the idea of precessing jets from a binary to explain the point symmetric structures observed in PNe is at least 30 years old (*11*).

Fleming 1 (PN G290.5+07.9, hereafter Fg 1) is a southern planetary nebula renowned for its spectacular set of bipolar jets delineated by a symmetric configuration of high speed knots (*12,13*). At an assumed distance of 2.4 kilo-parsec (*14*) they span about 2.8 parsec from tip to tip. The knots in the jets around Fg 1 follow a curved path distinguished by opposing pairs whose connecting lines intersect precisely with the position of the central star. The outermost knots are elongated and, because of their measured radial velocities of roughly 75 km/s with respect to the bulk motion of the nebula, were probably ejected about 16 000 years ago, whereas the innermost ones were possibly ejected some 9 or 10 000 years later (*12*). Based on its reported expansion velocity, one can also estimate the innermost nebula to be about 5 000 years old, assuming a constant expansion rate. Because of their appearance, suggestive of episodic ejections produced by a precessing source, Fg 1 became the archetype of a morphological sub-class of PNe named after their bipolar, rotating, episodic jets.

In 2011 we obtained a series of medium-resolution spectra of the central star of Fg 1 with FORS2 on ESO's Very Large Telescope (*15*; table S1). The velocities, measured from an average of the C IV lines at 581.1 and 581.2 nm, clearly show a periodic behavior (Fig. 1) that we attribute to the presence of a close companion. Using a Levenberg-Marquardt minimization method, we fitted a circular orbit with an orbital period of $1.1953 \pm 0.0002$ days and semi-amplitude of $87.65 \pm 1.68$ km/s, where the errors were estimated via a Bootstrap method. Treating the eccentricity as a free parameter, we find no evidence for a non-circular orbit ($e = 0.015 \pm 0.017$).

A non-local thermodynamic equilibrium analysis of the spectra shows that they can be fitted with a pure solar abundance hydrogen and helium atmosphere having an effective temperature, $T_{eff} = 80\ 000 \pm 15\ 000$ K, and a surface gravity, $\log g = 5.00 \pm 0.25$ cm/s$^2$ (*15*). A comparison with post-asymptotic giant branch (post-AGB) evolutionary tracks (*16*) implies an approximate mass of the primary star: $0.56^{+0.3}_{-0.04}$ solar masses.

The morphology of Fg 1 can be represented by a PN core that has the shape of a butterfly with its axis tilted at about 50 degrees to the line of sight (*17*). This is most likely also the orientation of the orbital plane (the main jet axis being perpendicular to it) and we can thus make use of this fact to estimate the mass of the secondary star using our measured orbital parameters. Assuming an inclination of $45 \pm 5$ degrees and using our value of the binary mass function, we derive a mass of the secondary between 0.7 and 1 solar masses. The companion must therefore be either an early K or G dwarf, or a more massive white dwarf. A main-sequence companion would, however, be irradiated by the hot primary – as seen for example in the 1.16-day binary system inside the Necklace nebula (*6*). We are able to discard any photometric variability above 0.05 mag (*15*), much smaller than would be expected for a 1.2-d orbital period and the range of effective temperature we consider (*18*). Moreover, the absence of any irradiated emission lines at all orbital phases (*15*) that would indicate the presence of a main-sequence companion provides additional proof that the system must be a double degenerate, that is, that the companion is a slightly more massive white dwarf that has either become too faint or whose spectral energy distribution peaks in the ultraviolet, and is no more detectable in the visible wavelength domain. For such a companion to provide the required number of ionizing photons above 54 eV to explain the observed ionization level of the nebula (which the observed central star cannot supply), its temperature needs to be greater than 120 kK. Such a high temperature would also make it undetectable in the visual spectrum. A more detailed analysis (*15*) shows that stars on post-AGB tracks with masses of 0.63 to 0.7 solar masses would meet these conditions. Thus, Fg 1 is most likely the latest addition to the very few examples of double degenerate systems found in PNe. For the two stars to be still inside a nebula and with the secondary being so hot, either the initial mass ratio was very close to unity and the two stars evolved towards the planetary nebula phase almost simultaneously, or the secondary has been reheated by accretion, just before the CE.

Our deep images of Fg 1, obtained with FORS2 in Hα+[N II], [O III] and [O II] filters, show both an envelope of shocked gas around the jets (Fig. 2) and a clear ring of low-ionization knots (Fig. 3). The characteristics of the fragmented jets in Fg 1 resemble the ballistic model predictions for jets from time-dependent sources. Three-dimensional gas-dynamical simulations show that the clumps are ejected in a bipolar outflow from a source in a circular orbit that has a precessing outflow axis (*19, 20*). At later times a mostly point-symmetric structure remains, in which the flow has developed flat-topped ends, very similar to what is seen in Fg 1. Another characteristic of these models is the presence of an envelope around the jets, similar to that seen in our images (Fig. 2). This envelope adds further weight to the argument that the precessing jets do in fact originate from a time-dependent source as prescribed by the models. Such precession is best explained by an accretion disk around a companion star and has been invoked to explain point-symmetric jets in other PNe (*21*). The presence of polar ejections older than the main body of the nebula – as recently found in a few systems (*6, 7, 22*) – indicates that mass transfer before the CE is most likely responsible for the jets. The extent of its S-shaped jets, a strong signature of precession, distinguishes Fg 1 from these other systems, and the reason may lie in the fact that it contains two degenerate objects.

The necessary accretion disc is likely to have formed around the secondary (*10*), most probably from material lost by the stellar wind (*23*). We conjecture that such a disc could have formed through "wind Roche-lobe overflow" (*24*), when the wind material fills the giant's Roche-lobe and is transferred to the companion through the inner Lagrangian point. This can dramatically increase the accretion rate with respect to normal wind accretion and has also been invoked to explain the current state of the symbiotic star SS Lep (*25*). Once the AGB star filled its Roche lobe, the mass transfer became unstable and a common envelope formed, shutting off the accretion disc and its associated jets, meanwhile forming the 5000-yr old inner nebula and shrinking the orbit of the binary system to the current observed value.

A double-degenerate central star would imply that Fg 1 must have gone through two successive mass transfer episodes. In the first one the system avoided the common envelope – as the remaining system must have been still wide enough to leave space for the secondary to expand to the AGB phase. In the second mass transfer event, described above, this was not the case. The existence of symbiotic and other peculiar red giant systems composed of a red giant and a white dwarf with orbital periods of several hundreds or thousands of days (*26*) shows that such stable episode of mass transfer is not uncommon, even though the details are still far from understood (*27*).

As with the jets in Fg1, its inner ring of knots (Fig. 3) is also thought to result from a binary interaction (*28, 29*). Such knotty rings are found in a variety of stars at the end of their evolution, one of the most spectacular cases being SN1987A (*30*). These rings are a distinctive feature of many PNe around close binaries (*31*), most notably in the Necklace (*7*). However, similar torus structures are observed in much wider binaries like symbiotic stars (e.g. *32*), as well as in several evolved massive stars (*33*). The examples provided by Fg 1, the Necklace, the other PNe with close binary central stars, and symbiotic Miras, point to a common mechanism linked to binary evolution for the formation of ring nebulae around many kinds of stars.

**References and Notes:**
1. B. Balick, A. Frank, *Ann. Rev. of Astron. Astrophys.* **40**, 439 (2002).
2. N. Soker, *Publ. Astron. Soc. Pacific* **118**, 260 (2006).
3. J. Nordhaus, E. G. Blackman, *Mon. Not. R. Astron. Soc.* **370**, 2004 (2006).
4 O. De Marco, *Publ. Astron. Soc. Pacific* **121**, 316 (2009).
5. B. Miszalski *et al.*, *Astron. Astrophys.* **496**, 813 (2009).
6. B. Miszalski *et al.*, *Mon. Not. R. Astron. Soc.* **413**, 1264 (2011).
7. R. L. M. Corradi *et al.*, *Mon. Not. R. Astron. Soc.* **410**, 1349 (2011).
8. D. Jones *et al.,* In Evolution of Compact Binaries, *ASPC* **447**, 165 (2011).
9. R. E. Pudritz *et al.*, In Protostars and Planets **V**, 277 (2007).
10. N. Soker, M. Livio, *Astrophys. J.* **421**, 219 (1994).
11. J. P. Phillips, N. K. Reay, *Astron. Astrophys.* **117**, 33 (1983).
12. J. A. Lopez, J. Meaburn, J. W. Palmer, *Astrophys. J. l* **415**, L135 (1993).
13. J. A. Lopez, M. Roth, M. Tapia, *Astron. Astrophys.* **267**, 194-198 (1993).
14. W. J. Maciel, *Astron. Astrophys. Suppl.* **55**, 253 (1984).
15. Additional information is available in the supplementary materials on *Science* Online.
16. D. Schoenberner, *Astrophys. J.* **272**, 708 (1983).
17. J. W. Palmer *et al.*, *Astron. Astrophys.* **307**, 225 (1996).
18. O. De Marco, T. C. Hillwig, A. J. Smith, *Astron. J.* **136**, 323 (2008).
19. J. A. Cliffe, A. Frank, M. Livio, T. W. Jones, *Astrophys. J.* **447**, L49 (1995).
20. A. C. Raga *et al., Astrophys. J.* 707, **L6** (2009).
21. L. F. Miranda *et al., Mon. Not. R. Astron. Soc.* **321**, 487 (2001).
22. D. L. Mitchell *et al., Mon. Not. R. Astron. Soc.* **374**, 1404 (2007).
23. T. Theuns, H. M. J. Boffin, A. Jorissen, *Mon. Not. R. Astron. Soc.* **280**, 1264 (1996).

24. S. Mohamed, P. Podsiadlowski, *Am. Inst. Phys. Conf. Ser.* **1314**, 51 (2010).
25. N. Blind *et al., Astron. Astrophys.* **536**, A55 (2011).
26. J. Mikolajewska, Baltic Astr. **21**, 1 (2012).
27. T. E. Woods *et al., Astrophys. J.* **744**, 12 (2012).
28. L. Sandquist *et al., Astrophys. J.* **500**, 909 (1998).
29. A. C. Raga *et al., Astron. Astrophys.* **489**, 1141 (2008).
30. S. Mattila *et al., Astrophys. J.* **717**, 1140 (2010).
31. B. Miszalski *et al., Astron. Astrophys.* **505**, 249 (2009).
32. M. Santander-García *et al., Astron. Astrophys.* **465**, 481 (2007).
33. N. Smith, J. Bally, J. Walawender, *Astron. J.* **134**, 846 (2007).

**Acknowledgments:** This paper uses data from ESO programmes 084.C-0508(A), 085.D-0629(A) and 087.D-0446(B), and includes observations made at the South African Astronomical Observatory (SAAO). The measured radial velocities and observed spectrum are provided in the Supplementary Materials. Observations obtained with ESO telescopes can also be obtained from the ESO science archive at http://archive.eso.org, using the above mentioned programme numbers. The SAAO photometric data are available from http://www.eso.org/~hboffin/Fg1/ .

BM thanks ESO for their hospitality and the opportunity to participate in their visitor program during January 2012. T.R. is supported by the German Aerospace Center (DLR) grant 05 OR 0806. This work was co-funded under the Marie Curie Actions of the European Commission (FP7-COFUND). The work of RLMC has been supported by the Spanish Ministry of Science and Innovation (MICINN) under grant AYA2007-66804. ADJ is supported by a FONDECYT "postdoctorado" fellowship under project number 3100098. ADJ is also partially supported by the Joint Committee ESO-Government of Chile.

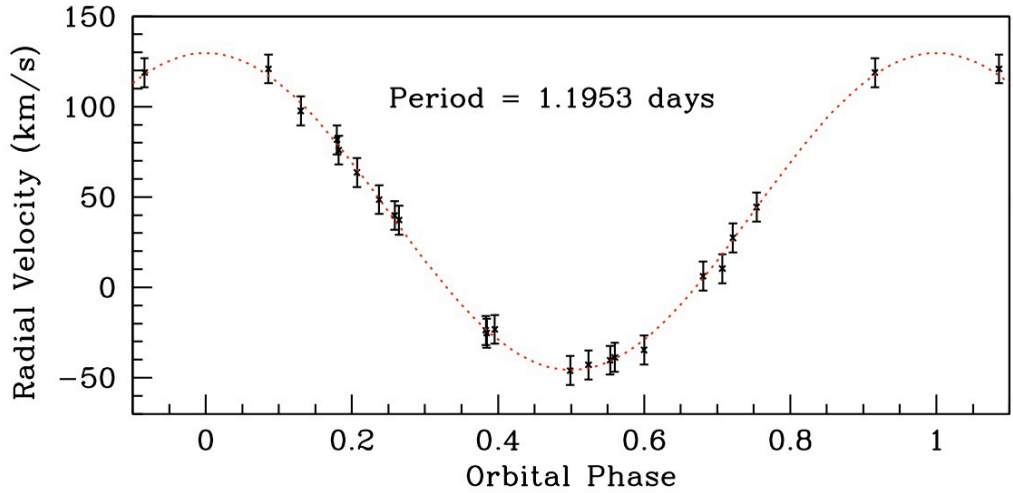

**Fig. 1.** Radial velocity curve of the central star of Fg 1. The measured radial velocities are folded on the orbital period of 1.1953 days and a sinusoidal curve is indicated for comparison.

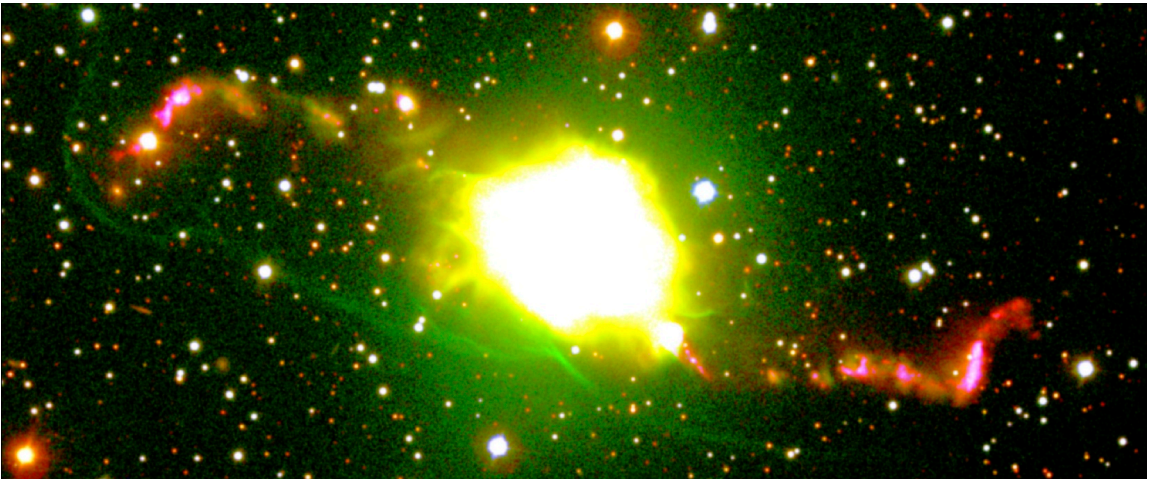

**Fig. 2.** FORS2 colour-composite image of Fg 1 showing the [O III] envelope around the bipolar jets. The 5.5' x 2.3' image is based on individual images obtained through Hα+[N II] (red), [O III] (green) and [O II] (blue) filters. The orientation is the same as in Fig. 3.

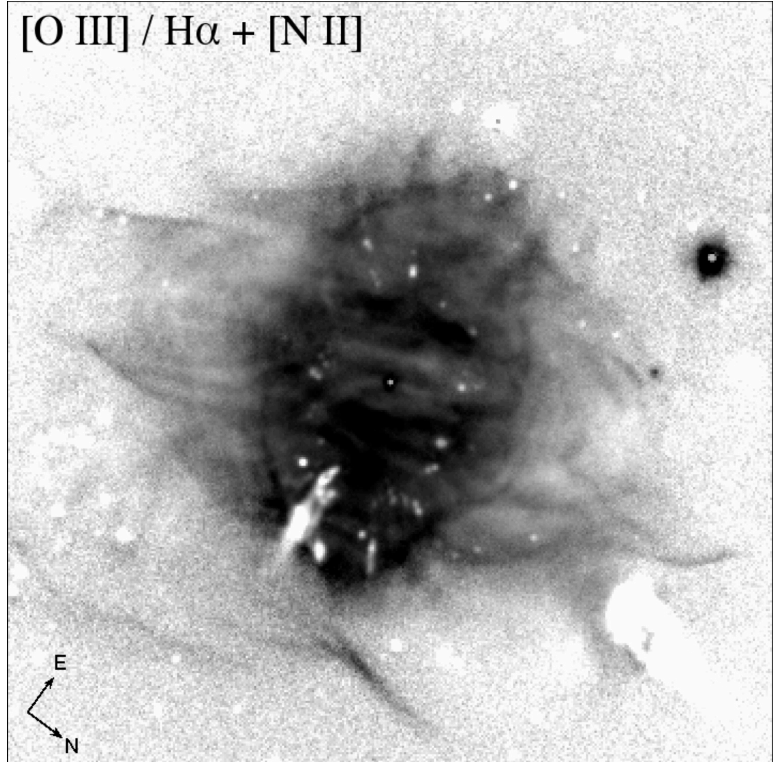

**Fig. 3.** The central nebula of Fg 1: this 90" x 90" image shows the ratio between the [O III] and Hα+[N II] FORS2 images. Black corresponds to high ratios ([OIII] bright), while white corresponds to small ratios (Halpha+[NII] bright). The central star (CSPN) and the low-ionization knots (forming an approximate ring roughly 35" x 64" in size) are highlighted.

**Table 1. Characteristics of Fg 1**

| Right ascension | 11 28 36.2 |
|---|---|
| Declination | -52 56 03 |
| Orbital Period, $P$ | 1.1953 ± 0.0002 days |
| To (MJD) | 55671.556 ± 0.018 days |
| Eccentricity, $e$ | 0 (fixed) |
| Radial Velocity semi-amplitude, $K$ | 87.65 ± 1.68 km/s |
| Mass function, $f(m)$ | 0.084 ± 0.005 solar mass |